\documentclass{aa}
\usepackage{psfig}
\begin{document}

   \thesaurus{06     
              (08.09.2 RX~J1141.3-6410;  
	       08.14.2;  
               08.13.1;  
	       02.01.2;  
	       02.16.2;  
	       13.25.5)} 
\title{Polarimetry and spectroscopy of the polar
RX~J1141.3-6410\thanks{This work was based on observations made at
Laborat\'orio Nacional de Astrof\'\i sica/CNPq/MCT, Brazil.}}

\author{C.~V.~Rodrigues\inst{1,2}
\and
D. Cieslinski\inst{2,1}
\and
J. E. Steiner\inst{2,3}}

\offprints{C. V. Rodrigues, e-mail: claudia@das.inpe.br}

\institute{Div. Astrof\'\i sica,
	   Instituto Nacional de Pesquisas Espaciais/MCT,
	   Caixa Postal 515,
	   12201-970, S\~ao Jos\'e dos Campos, SP, Brazil
\and
Depto. de Astronomia,
Instituto Astron\^omico e Geof\'\i sico/USP,
Caixa Postal 9638,
01065-970, S\~ao Paulo, SP, Brazil
\and
Laborat\'orio Nacional de Astrof\'\i sica/CNPq/MCT,
Av. Estados Unidos, 154,
37500-000, Itajub\'a, MG, Brazil
}

   \date{Received <date>; accepted <date>}

\titlerunning{Polarimetry and spectroscopy of the polar
RX~J1141.3-6410}
\authorrunning{Rodrigues et al.}
   \maketitle

   \begin{abstract}

We present the first optical polarimetric measure\-ments of
RX~J1141.3-6410 which confirm that star as a polar. The circular
polarization varies between 0 and 13\% with the orbital period.
H$\alpha$ spectroscopy shows that this line is formed by, at least, two
components: a broad and a narrow one. The phase of maximum redshift of
the broad component is shifted by 0.5 with the phase of maximum
circular polarization which is not usual for this class of stars.  We
suggest a geometrical configuration for the system which could explain
the main features of the polarimetric and spectroscopic data.

   \end{abstract}

\section{Introduction}
Cataclysmic variables (CVs) are binaries consisting of a red main
sequence star and a white dwarf (primary). The secondary fills its
Roche lobe and mass is transfered to the white dwarf (WD). This process
usually forms an accretion disk. However, some CVs have a magnetic
field intense enough to prevent the disk formation. In this case, the
matter falls onto the WD near the magnetic pole forming an accretion
column.  Another important consequence of the high strength of the
magnetic field is the synchronization of the white dwarf rotation with
the orbital movement.  These stars are denominated polars and their
prototype is AM Her. Reviews can be found in Cropper (1990) and Warner
(1995).  The polars can be distinguished from intermediate polars
(non-synchronized magnetic systems) by two important features: high
circular polarization and strong soft X-ray emission. The former is
caused by cyclotron emission in the column. The emission at high
frequencies is produced by the shock formed close to the WD surface.

The emission lines in polars are thought to be formed along
the trajectory of the material from the secondary to the white
dwarf (see review by Mukai 1988). This material leaves the secondary at
the inner Lagrangian point ($L_1$) and follows its ballistic trajectory
in the orbital plane (horizontal stream) down to the coupling region. In
this region the magnetic pressure becomes higher than the ram pressure
and the material starts to follow the magnetic lines (accretion
stream). The radiation produced near the white dwarf can be
reprocessed on the secondary surface contributing to the emission
lines.

The emission lines of polars can be formed by up to four components
(Rosen et al. 1987, e.g.). In general, at least two components are
seen: a broad base component and a narrow peak component.  The
accretion stream is responsible for the broad component.  The narrow
one is formed nearer the secondary. In some systems, this emission
seems to be formed on the secondary surface itself (Liebert \& Stockman
1985). In others, there is evidence that the horizontal stream produces
such emission (Mukai 1988).

Recently, Doppler tomography of polars has improved the understanding
of the emission lines (Diaz \& Stei\-ner 1994; Schwope et al. 1997;
\v{S}imi\'c et al. 1998).  An important fraction of the emission seems
to be produced near the secondary.  The bulk of the broad component is
formed near the coupling region.  These maps do not show an
important emission from the region of high velocities near the white
dwarf.

Many sources identified by the ROSAT satellite have been shown to be
CVs and, more specifically, polars.  Motch et al.  (1996) discovered 7
new CVs and suggested that two of them could be synchronized magnetic
systems based on the strength of their emission lines. RX~J1141.3-6410
is one of these systems.  It is associated with a $\approx$16.5~mag
star having a strong \ion{He}{ii} $\lambda$4686 emission line.
Recently, Cieslinski \& Steiner (1997) have found a photometric period
(P = 0.131\,517~d) and a light curve consistent with the suggestion of
RX~J1141.3-6410 being a polar. However, until now, no polarimetric
measurement has been made in order to confirm its magnetic nature.

In this work, we present our optical polarimetric measurements and
time-resolved spectral data in the region of H$\alpha$ for
RX~J1141.3-6410.  Some modeling of the intensity and polarization has
been made. We suggest a possible geometrical configuration for
RX~J1141.3-6410 and the main regions of line formation based on the
polarization models and the spectroscopic data.

\section{Observations and data reduction}
\subsection{Polarimetric acquisition and reduction}

The polarimetric observations were made during two nights (March 04-05,
1997) with the 1.60~m telescope of the {\it Laborat\'orio Nacional de
Astrof\'\i sica} (LNA), Brazil. We used a CCD camera modified by a
polarimetric modulus (Magalh\~aes et al. 1996). All 
measurements were made using a $R_C$ filter.

The polarimetric modulus consists of a fixed analyzer (calcite prism),
a $\lambda$/4 retarder waveplate and a filter wheel. The retarder plate
is rotated with 22\fdg5 steps. Therefore, a polarization measurement
consists of (a minimum of) eight integrations in subsequent retarder
orientations. The calcite block separates the extraordinary and
ordinary beams by 12\arcsec. This division eliminates any sky
polarization (Piirola 1973; Magalh\~aes et al.  1996). The $\lambda$/4
retarder allows us to measure the circular and linear polarization
simultaneously.

Our main goal was to measure the circular polarization (V) with an
error of the order of 1\% and a time resolution enough to check a
variability locked with the photometric modulation. Therefore, we chose
a time integration of 90 s for each individual image. In this way, a V
point spans 8 $\times$ 90~s (720~s) plus the dead time. We would
usually group the images in sequences of 8 images with no overlap
(images 1 through 8, images 9 through 16 and so on). However, in order
to improve the temporal resolution we have grouped the images with
overlap (images 1 through 8, images 2 through 9 and so on).  In this
way, we have an time interval between two points of typically 140~s
(90~s + dead time).  These data also enable us to perform differential
photometry using comparison stars in the field.

The basic reduction of the polarimetric data is identical to the
photometric one. This step was performed using standard
IRAF\footnote{IRAF is distributed by National Optical Astronomy
Observatories, which is operated by the Association of Universities for
Research in Astronomy, Inc., under contract with the National Science
Foundation.} routines for image correction and photometric analysis.
The derived counts were the input for a FORTRAN code that calculates
the polarization (Ma\-galh\~aes et al. 1996; Magalh\~aes et al. 1984).
However, in those works, one cannot find the solution for the Stokes
parameters using a $\lambda/4$ plate, which is presented below. 

The intensities of the two images in the CCD, ordinary ($I_o$) and
extraordinary ($I_e$) ones, are related to the Stokes parameters (I, q,
u, v) of the incident beam by (Serkowski 1974):

\[
2 I_{o,e} = I \pm q \cos^2 2\theta \pm u \sin 2\theta \cos2\theta \mp v 
\sin2\theta,
\]

\noindent where $\theta$ is the waveplate position angle. The upper and lower signals refer to each of the two images. 

The normalized Stokes parameters (Q=q/I, U=u/I and V=v/I) can be
obtained using the method of Magalh\~aes et al. (1984) applied to a
$\lambda$/4 retarder. Defining the quantity $z_i$ for each retarder
position, i:

\[
 z_i = \frac{I_{e,i} - I_{o,i}}{I_{e,i} + I_{o,i}},
\]

\noindent we obtain the following expressions for the Stokes parameters
(assuming eight retarder positions):

\begin{eqnarray*}
Q & = & \frac{1}{3} \sum z_i \cos^2 2 \theta_i, \\
U & = & \sum z_i \sin 2 \theta_i\cos 2\theta_i, \\
V & = & -\frac{1}{4} \sum z_i \sin 2 \theta_i,
\end{eqnarray*}

\noindent where $\theta_i$ is the retarder position angle.

The first position of the retarder did not correspond to zero degree.
But, it was possible to determine its alignment by minimizing the
errors in V. We are confident that this procedure gives us a good
estimate of the zero point by the results obtained for the observed
standard stars.  However, we could not solve an indetermination of
180$\degr$. Because of that we could only measure the V modulus, not
its signal.

One of the images of RX~J1141.3-6410 was contaminated by the
superposition of a much fainter neighbour star.  We did the photometry
of the other image of the superposed star, which was isolated, and
these counts were subtracted from those of RX~J1141.3-6410.  We have
performed the reduction using different combinations of apertures for
RX~J1141.3-6410 and the superposed star and also using no correction at
all.  The circular polarization was practically not affected by this
procedure.  On the other hand, the linear polarization was sensitive to
the parameters used in the correction.  Therefore, the linear
polarization results must be taken with care.

The results of circular polarimetry and differential photometry are
presented in Fig. \ref{fig_circ}.  The average error in V is 0.82\%.
The errorbars of the linear polarization prevent us to detect any
polarization less than 4\%. Within this limit, no linear polarization
was detected in any orbital phase.

\begin{figure}[htpb]
\vspace{1cm}
\caption{Differential photometry and circular polarization of
RX~J1141.3-6410.  The abscissa corresponds to the orbital phase using
the new ephemeris of Sec. \ref{fotometria}.  The lines represent
the model of Wickramasinghe \& Meggitt (1985) with an electronic
temperature of 10 keV and $\Lambda$ equals to $10^5$ (see Sec.
\ref{pol_disc}). The full line corresponds to a model with: $i=5\degr$,
$\beta=87\degr$. The dotted line corresponds to a model with:
$i=85\degr$, $\beta=8\degr$.  Both models have a magnetic field
strength of 30MG}
\label{fig_circ}
\end{figure}

\subsection{Orbital ephemeris}
\label{fotometria}

The light curve shown in the previous section can be used to refine the
orbital ephemeris of Cieslinski \& Steiner (1997). Additional $R_C$
photometry were obtained on 1997 April 15-16-19 with the Boller \&
Chivens 60~cm telescope at LNA, using a CCD camera. The new ephemeris
for the maximum photometric optical flux is:

$$ T_{max}(HJD) = 2\,450\,356.46655(\pm3) + 0.131\,517\,8(\pm3)E .$$

All orbital phases quoted in this paper refer to the above ephemeris.

\subsection{Spectroscopic data}
\label{spec_data}

We obtained 30 spectra with 10 minutes of integration time on April 14,
1997 with the 1.60~m telescope of the LNA. We used the Cassegrain
spectrograph with a 1200 line mm$^{-1}$ grating and a CCD detector.
This configuration gave a spectral resolution of about 2~\AA\ and a
covered wavelength region between 6100 and 6970~\AA.   We used a
200$\mu$m slit which corresponds to 2\arcsec\ in the sky.  This was
approximately the seeing in that night.  A GG385 order blocking filter
was also used.  The reduction was made using the standard routines of
the IRAF package. The spectra were flux calibrated using the
spectrophotometric standard stars from Stone \& Baldwin (1983) and
Taylor (1984).  Fig. \ref{spectra} shows the spectra binned in 10
orbital phases (see discussion below).  The H$\alpha$ and \ion{He}{i}
$\lambda 6678$ lines are present.

Albeit the small
range in wavelength, we have searched for some indicative of a
cyclotron component in the spectra.  For this we have averaged the
spectra corresponding to the faint and bright photometric phases
separately.  Within the wavelength range and the signal to noise ratio
of our spectra, no difference in the continua could be noted.

\begin{figure}[htpb]
\vspace{1cm}
\caption{Spectra of RX~J1141.3-6410 binned in 10 orbital phases}
\label{spectra}
\end{figure}

In order to check the existence of periodicity in the spectroscopic
data, we have calculated the radial velocity of both lines using the
Double Gaussian method described by Schneider \& Young (1980) and
Shafter (1985). This method measures the radial velocity of the line
wings. Therefore, it provides us with the radial velocity of the broad
base component. The Double Gaussian method requires two adjustable
parameters:  the Gaussian width, $\sigma$, and the distance between the
curves, $2a$.  They have been chosen by the minimization of the
$\sigma_K/K$, where K is the semi-amplitude of the radial velocity
curve. The K-amplitude has been obtained from the fit of a sinusoidal
model.  For the H$\alpha$ line, we have found:  $\sigma$ = 12~\AA\ and
$a$ = 8~\AA. The Discrete Fourier Transform method applied to the
derived radial velocities has given a spectroscopic period of 0.1290
$\pm$ 0.0022~d. This value is completely consistent with the
photometric one within one sigma uncertainty. The result for the
\ion{He}{i} $\lambda 6678$  line is also consistent with the
photometric period,  but the errorbar is higher.  Based on the
agreement between the photometric and spectroscopic periods, we have
combined the original data in ten spectra according to the photometric
orbital phase (see Fig. \ref{spectra}).

An inspection of the spectra shows that the emission lines may be
formed by more than one component, which is common in this type of
star.  So we have deconvolved the ${\rm H}\alpha$ line using two
Gaussian functions.  The \ion{He}{i} $\lambda 6678$\  line was too
noisy preventing this analysis (see Fig. \ref{spectra}).  The two
Gaussian fit of the spectral profile has been done using the {\it
splot} routine of IRAF.  In order to constrain the parameters to be
adjusted, we have again applied the Double Gaussian method to the
combined spectra and the obtained central wavelengths of the broad
component have been assumed fixed. The best fit has been found for:
$\sigma$ = 10~\AA\ and $a$ = 19~\AA . The increase in the separation of
the two Gaussian for the combined spectra reflects the higher
signal/noise which has allowed us to detect the line further from the
peak.

The results for the radial velocities, flux, FWHM and equivalent width
are shown in Fig. \ref{two_gauss}. The widths (FWHM) have not been
deconvolved from the instrumental resolution which is around
90~km~s$^{-1}$.  Assuming the radial velocity of the broad component as
a free parameter, the flux, equivalent width and FWHM do not
practically change.  However, the departure of the radial velocity
relative to a sinusoidal curve increases: the maximum redshift tends to
occur at smaller phases ($\approx$0.35) and the maximum blueshift
remains at the same location.  The radial velocity curves of Fig.
\ref{two_gauss} have been fitted using a standard sinusoidal model,
whose results are shown in Tab.  \ref{rad_vel} and Fig.
\ref{two_gauss}.  Using the individual 30 spectra, we have obtained
similar values for the broad component.  For instance, the K-amplitude
is K=110.5 $\pm$ 5.8.

\begin{figure}[htpb]
\vspace{1cm}
\caption{Radial velocities, flux, FWHM and equivalent widths of the
broad and narrow components of the ${\rm H}\alpha$ line of
RX J1141.3-6410 obtained using a two Gaussian fit.
The curves in the radial velocity panel correspond to the fitted
sinusoidal model (see Tab. \ref{rad_vel})}
\label{two_gauss}
\end{figure}

\begin{table}
\tabcolsep 50 pt
\caption{Radial velocities parameters of the broad and narrow components
of the ${\rm H}\alpha$ line of RX~J1141.3-6410}
\label{rad_vel}
\[
\begin{array}{ccc}
\hline
\noalign{\smallskip}
\ \ \ {\rm Parameter} \ \ \ &  \ \ \ {\rm Broad\ Comp.}\ \ \ & \ \ \ 
{\rm Narrow\ Comp.}\ \ \  \\
\noalign{\smallskip}
\hline
\noalign{\smallskip}
 K~(\mathrm{km~s^{-1}}) & 120.2 \pm 8.2 & 57.1 \pm 7.8 \\
\gamma~(\mathrm{km~s^{-1}}) & 69.5 \pm 5.9 & 14.2 \pm 5.5 \\
\phi_o^{\rm a} & 0.444 \pm 0.011 & 0.783 \pm 0.022 \\
\noalign{\smallskip}
\hline
\end{array}
\]
\begin{list}{}{}
\item[$^{\rm a}$] This phase corresponds to the maximum redshift of
the component
\end{list}
\end{table}

\section{Discussion}

\subsection{What the polarization of RX~J1141.3-6410 can tell us}
\label{pol_disc}

Figure \ref{fig_circ} shows the flux (in arbitrary units) and the
circular polarization (modulus) of RX~J1141.3-6410. A high degree of
circular polarization is present in a large fraction of its period with
a maximum value of approximately 13\%. This confirms the classification
of RX~J1141.3-6410 as a polar.  This figure also shows a clear
correlation between flux and polarization. The phases of null
polarization coincide with the photometric minimum indicating that in
these phases the accretion region is out of the line of sight.
Whatever the signal of V is, no change in the signal is observed
(because V crosses the zero axis once).  This indicates that in this
wavelength region we probably see only one pole or regions of same
polarity.

We have fitted our results on flux and circular polarization using the
models of Wickramasinghe \& Meggitt (1985, hereafter WM85) for a
point-like accretion region. This region must be extended (Mukai 1988),
but a point-like model can provide us with some insight into the
geometrical and physical configuration of RX~J1141.3-6410.  In order to
do the fits, we have assumed two components: (1) a component which is
constant along the orbital period and whose polarization is null; and
(2) a component coming from the accretion region whose intensity and
polarization are assumed to be represented by the WM85 models.  We have
normalized the WM85 intensity values to fit the maximum value of the
second component.  With the above assumptions, the circular
polarization, V, can be also determined without any extra
normalization. So the level and the shape of the circular polarization
can be fitted.

The phase duration of constant flux and polarization (V = 0\%) can be
identified with the phase where the accretion region is occulted from
the observer. This phase interval, $\Delta\phi$, depends only on the
geometry of the system and defines a relation between the orbital
inclination, $i$, and the colatitude of the magnetic field, $\beta$
(Bailey \& Axon 1981):

\begin{equation}
\cos\pi\Delta\phi = \frac{1}{\tan i \tan\beta}.
\label{ba81}
\end{equation}

From Fig.  \ref{fig_circ}, we see that this phase interval is
around 0.30. So hereafter we will define the geometry only by the
inclination, since the above relation fixes the value of $\beta$.

We could find reasonable fits to all combinations of electron
temperature and size parameter, $\Lambda$, presented by WM85. Examples
are shown in Fig.  \ref{fig_circ}.  It should be noted, however, that V
is better adjusted than the flux (see Fig. \ref{fig_circ}). This is
consistent with the results of Piirola et al. (1990) who used the
calculations of WM85 to construct a model for an extended emitting
region. Their results show that the flux is more modified than the
circular polarization relative to the point-like model (see their Fig.
5).  For an extended region the flux tends to have a smaller flat top
and consequently a larger phase interval of increasing (and decreasing)
flux. Therefore the changes caused by the extension of the emitting
region and possible gradients in the physical conditions would improve
the agreement between the model and our data.

Considering only the curve shapes, we could get a hint about
the orbital inclination of RX~J1141.3-6410 (and consequently $\beta$).
The model indicates two possible solutions:  small inclinations (around
5$\degr$) or large inclinations ($\approx$ 85$\degr$). For intermediate
values, the polarization and flux curves present peaks or are not
correlated.  This happens because the intensity and polarization have
opposite behaviours as $\alpha$ varies, where $\alpha$ is the angle
between the line of sight and the dipole axis of the magnetic field
(WM85).  This implies that a correlation between intensity and
polarization could only be achieved if $\alpha$ does not change so
much. As we have a phase where $\alpha$ must be around 90$\degr$ (the
phase when there is a transition from a visible to a hidden column),
$\alpha$ must be restricted around this value.  Using Eq. \ref{ba81},
the colatitude of the magnetic field axis is about 0$\degr$ for
inclinations near 90$\degr$. If the system is seen face on, $\beta$
must be near 90$\degr$. The numerical solutions for the intensity and
circular polarization are exactly the same if one considers ($\pi-i$)
and ($\pi-\beta$).

Assuming a fixed geometry and one of the WM85 models, the polarization
modulus depends basically on the magnetic field strength.  We have
found values restricted to the range between 15 and
40 MG.

There is no evidence of eclipses in the light curve of
RX~J1141.3-6410.  This can impose a limit to its inclination angle.
First, let's estimate the maximum inclination, $i_{max}$, that allows a
non-eclipsing system.  It can be seen that:

\begin{equation}
\sin(\frac{\pi}{2}-i_{max}) = \frac{R_2}{a},
\end{equation}

\noindent where $R_2$ is the secondary radius and $a$ is the distance
between the stars. The secondary radius can be approximated by the size
of its Roche lobe. So, one could use the expressions of Paczynski
(1971) for $R_2/a$ in order to obtain $i_{max}$. These relations depend
on the mass ratio of the system, q.

Using the relation of Patterson (1984) between the mass of the
secondary, $M_2$, and the orbital period in cataclysmic variables, we
obtain $M_2 = 0.28M_\odot$. The average primary mass of a magnetic
cataclysmic variable is $ 0.79 \pm 0.11M_\odot$ (Webbink 1990).  This
give us an approximate value to q of 0.36.  This corresponds to
$i_{max}$ equal to 73$\degr$. Assuming values of the primary between
0.6 and 1.4$M_\odot$, the $i_{max}$ range is 71\fdg 6 to 75\fdg 3.  The
WM85 models with an inclination of 75$\degr$ produces a circular
polarization quite different from the observed (much more peaked). So
the models of the cyclotron component tend to favour very low values
for the inclination.

\subsection{Emission lines in RX~J1141.3-6410}

Like the polarization and optical flux, the orbital variation of the
emission line components depends on the geometrical and physical
configuration of RX~J1141.3-6410. Below, we present some discussion on
the spectroscopic data presented in Sec. \ref{spec_data}.

The narrow component of polars can be associated with the irradiated
surface of the secondary or the stream.  In RX~J1141.3-6410, we favour
the interpretation that the narrow component is due to the horizontal
stream. The main argument is the relatively high (deconvolved) width of
this component which can reach more than 200 km~s$^{-1}$ (Mukai 1988).
The orbital dependence of the flux can also give us some information
about the region of line formation.  The flux of a component originated
on the secondary surface has only one peak during the orbital period
(e.g., Schwope et al. 1997; Beuermann \& Thomas 1990).  It occurs at
the superior conjunction of the secondary when the largest fraction of
its irradiated surface is exposed to the observer.  On the other hand,
an optically thick stream has its maximum flux when it is seen sideways
(Mukai 1988).  This produces two peaks during one orbital period.  The
flux of the narrow component is not very accurately determined, but
there is a small evidence for two peaks along the period. So that could
also indicate that the stream is the main contributor to the narrow
component.  The K-amplitude of the narrow component radial velocity is
small ($\approx$ 60 km s$^{-1}$).  Any movement in the orbital plane
would have small radial velocities, if the system is seen near pole on
- as the absence of eclipses seems to indicate. Therefore the
K-amplitude could not help in disantagling the origin of the narrow
component.

The understanding of the geometrical configuration of RX~J1141.3-6410
would be improved if we had some information about the secondary
phasing. This could be  directly obtained through eclipses, absorption
lines or ellipsoidal variations.  Unfortunately, none of them is
available for this star.  The blue-to-red phasing of the radial
velocities of lines formed in the horizontal stream usually presents a
small offset relative to the inferior conjunction of the secondary
between -0.05 and -0.3 (Mukai 1988). If the narrow component is indeed
produced in the horizontal stream, the inferior conjunction of the
secondary would occur near phase 0.6. This puts the accretion
region on the WD face opposite to the secondary. Most polars
seem to have the active pole facing the secondary (Cropper
1990), contrary to what is seen is RX~J1141.3-6410, if the
above suppositions are correct.

About the narrow component, we could also say that if it is
produced in the orbital plane of the system, which is valid for the
stream or the secondary surface, the gamma velocity, $\gamma$, can be
associated with the systemic velocity of the binary.

The K-amplitude of the broad component in
polars can reach 1000 km~s$^{-1}$ (Warner 1995). This makes the values
found for RX~J1141.3-6410 relatively small ($\approx$120~km~s$^{-1}$).
This result is consistent with the polarization models, since they
imply that we are seeing the column edge on ($\alpha$ is approximately
90$\degr$). So the material falling down to the white dwarf has its
larger velocity component perpendicular to the line of sight.

The broad component flux does not seem to be occulted (as the cyclotron
component is), this can be an indication that it is formed at higher
distances from the white dwarf than the cyclotron component. This
supposition is consistent with the behaviour of its the equivalent
width. It has one maximum in the phase interval where the (polarized)
cyclotron component is occulted by the white dwarf.  The flux of the
broad component has two maxima in one period. They occur near the
maximum red and blueshift. In the maximum redshift phase (minimum
photometric flux) there seems to be a slightly smaller flux than in the
blueshift peak.

The width of the broad component presents two peaks during the orbital
period. It is consistent with the model by Ferrario et al.  (1989) of the
radial velocity and  width of the broad component.  For some models, the
width has two peaks which are not necessarily coincident with the
maximum projected velocities (see their Fig. 6).  More than that, the
difference between maximum radial velocity and maximum width increases
with the distance of the emission forming region from the WD.

The broad component of RX~J1141.3-6410 has its maximum redshift
approximately half a period after the maximum polarization. This is not
usual for polars, which commonly present the maximum redshift at the
same phase of the maximum circular polarization (Liebert \& Stockman
1985). The last happens when the accretion column is aligned with the
line of sight. Because of that, Liebert \& Stockman associated the
broad component  with the accretion funnel. However, Doppler tomography
of polars (Diaz \& Steiner 1994; Schwope et al. 1997; \v{S}imi\'c et
al.  1998) shows that the bulk of the broad component seems to be
produced near the coupling region. We suggest that the broad component
could possibly come from the coupling region. As the polarization must
come from a region nearer the WD (the falling region), this
configuration could explain the blueshift in the broad component at the
same phase of maximum circular polarization. However, we probably see
only a small component of the velocity, so these conclusions must
be taken with care.

If the broad component is indeed formed in the coupling region of
RX~J1141.3-6410, it is equivalent to say that it is formed
in the rising part of the accretion stream. This means that $V_z$ of
this component is positive. In order to get a $\gamma$ velocity
positive the inclination of the system must be greater than
90\degr. For RX~J1141.3-6410, particularly, this means around
180\degr.

\section{Conclusion}

The high level of circular polarization observed in RX~J1141.3-6410
confirms this star as a polar. Within our data precision, no peak in
linear polarization was observed. The ${\rm H}\alpha$ line of
RX~J1141.3-6410 shows two components. The maximum blueshift of the
broader component is locked with the maximum circular polarization,
contrary to most polars.

We suggest that the system is seen near face on and that the magnetic
field axis lies near the orbital plane. This
configuration is able to explain the main features of RX~J1141.3-6410
data.
The narrow component seems to be produced in the horizontal stream,
while the broad component may have its origin in the coupling region.

\begin{acknowledgements}
We are very thankful to the referee's suggestions which helped to
improve the paper.  We acknowledge Dr. M. P. Diaz by his careful
reading of the manuscript.  We are grateful to M. G. Pereira for
sharing telescope time.  CVR and DC would also like to acknowledge Dr.
F. J.  Jablonski for the discussions and his help with photometric
analysis. Finally, we are also thankful to A. Pereyra and A. M. Magalh\~aes for
providing their IRAF routines which facilitate the polarimetric
reduction.

\end{acknowledgements}

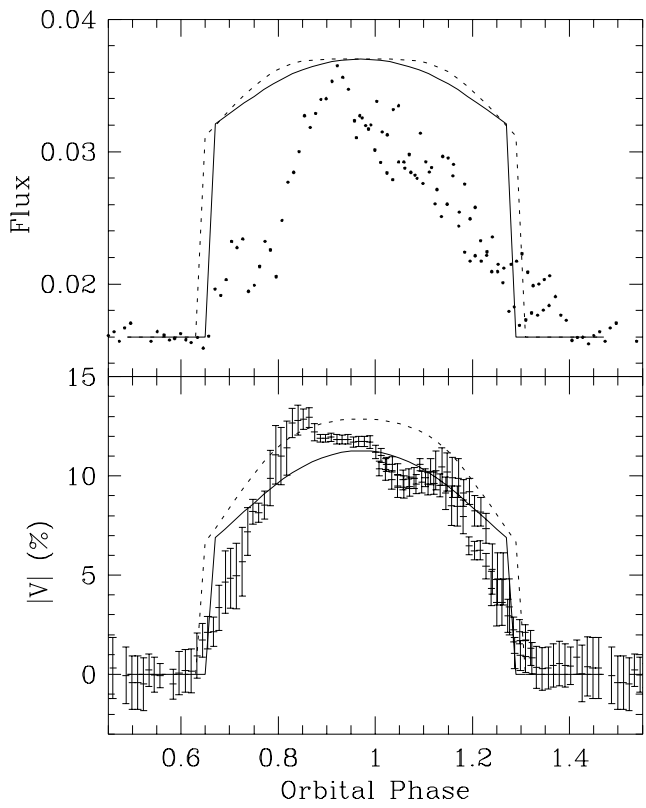
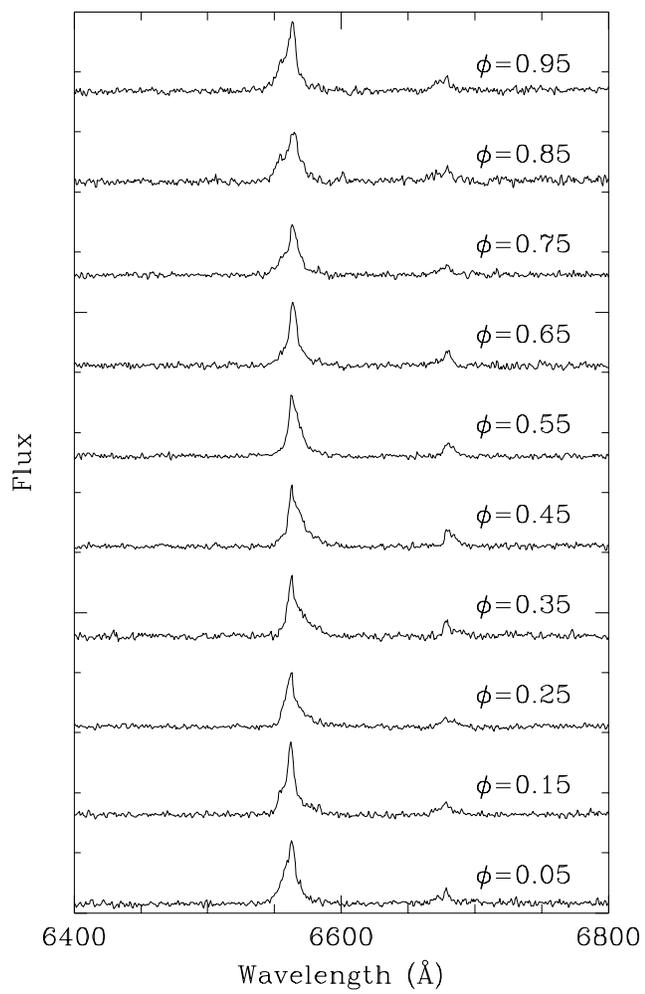
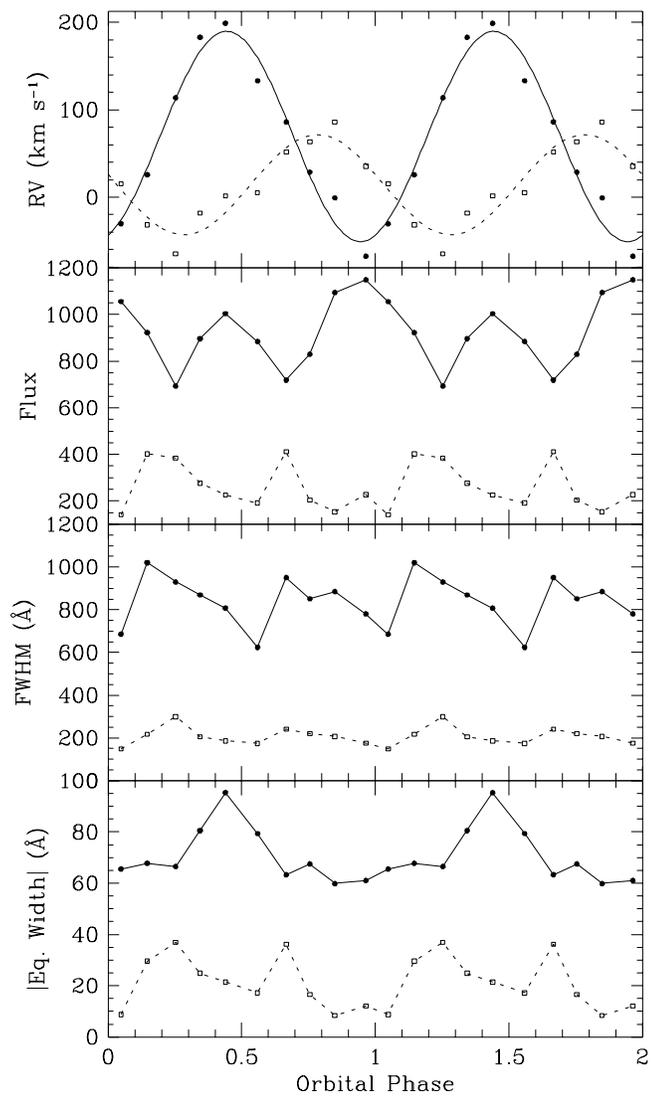
\end{document}